\documentclass[12pt,journal,onecolumn,draftclsnofoot]{IEEEtran}
\usepackage{amsmath,amssymb}
\usepackage{cite}
\usepackage{bm}
\usepackage{graphicx}
\usepackage{placeins}
\usepackage{booktabs}
\usepackage{tabularx}
\usepackage[letterpaper,margin=0.625in]{geometry}
\usepackage{setspace}
\usepackage{etoolbox}
\makeatletter
\def\@IEEEabskeysecsize{\normalsize}
\apptocmd{\thebibliography}{\normalsize\setstretch{1.30}}{}{}
\makeatother
\AtBeginDocument{%
  \setlength{\abovedisplayskip}{6pt plus 2pt minus 2pt}%
  \setlength{\belowdisplayskip}{6pt plus 2pt minus 2pt}%
  \setlength{\abovedisplayshortskip}{3pt plus 2pt}%
  \setlength{\belowdisplayshortskip}{4pt plus 2pt minus 2pt}%
  \setlength{\jot}{3pt}%
}

\begin{document}
	
\title{\fontsize{18}{24}\selectfont Energy-Efficient Risk-Constrained Adaptive Semantic Communications Under Uncertain Adversarial Threats}
	
	\author{
		Maedeh~Fallahreyhani,~\IEEEmembership{Student Member, IEEE,}
		Paeiz~Azmi,~\IEEEmembership{Senior Member, IEEE,}
		Nader~Mokari,~\IEEEmembership{Senior Member, IEEE,}
		M.~Reza~Abedi,~\IEEEmembership{Member, IEEE}
		\thanks{M. Fallahreyhani, P. Azmi, N. Mokari, and M. R. Abedi are
			with the Department of Electrical Engineering, Tarbiat Modares University,
			Tehran, Iran (e-mail: m.fallahreyhani@modares.ac.ir;
			pazmi@modares.ac.ir; nader.mokari@modares.ac.ir;
			mohammadreza\_abedi@modares.ac.ir).}
	}
	
	\markboth{Submitted for review to IEEE Transactions on Green Communications and Networking (TGCN)}%
	{Submitted for review to IEEE Transactions on Green Communications and Networking (TGCN)}
	
	\maketitle

\begin{center}
\small\itshape
This manuscript has been submitted for review to\\
IEEE Transactions on Green Communications and Networking (TGCN).
\end{center}
	
	\begin{abstract}
Semantic communication systems based on deep joint source--channel coding
can provide task-oriented wireless transmission, but adversarial robustness
may require substantial radio and receiver-side defensive-computation energy.
This paper develops an energy-efficient, risk-constrained semantic
communication framework under source-domain, over-the-air (OTA), and
potentially composite adversarial conditions. Long-term total energy,
including communication and diffusion-purification energy, is minimized
subject to average semantic-loss, trajectory-level tail-risk, and end-to-end
latency constraints. A recurrent primal--dual proximal policy optimization
controller jointly adapts transmit power, semantic/channel bandwidth ratio,
and purification complexity using causal channel observations, semantic
confidence, imperfect attack-belief estimates, and recurrent history. A
conditional value-at-risk (CVaR) constraint explicitly limits severe semantic
failures. All reported performance results are obtained from end-to-end
experiments on the CIFAR-100 dataset, including separate source-domain and OTA
adversarial evaluations. At the strongest evaluated
source-domain perturbation, $\epsilon=12/255$, the implemented controller
achieves $60\%$ robust Top-1 accuracy with empirical
$\mathrm{CVaR}_{0.9}\approx0.38$ and normalized total energy $\approx0.756$,
about $10\%$ below diffusion-only defense. At the strongest evaluated OTA
condition, $\Gamma_A=-5$ dB, it achieves $52\%$ robust Top-1 accuracy with
normalized total energy $\approx0.755$; its empirical
$\mathrm{CVaR}_{0.9}\approx0.42$ slightly exceeds the prescribed $0.40$ risk
threshold, explicitly exposing the robustness--energy--risk trade-off under
severe attack. A common accounting boundary links radio energy, purification
energy, and latency to the same realized control actions.
\end{abstract}
	
	
	
	\section{Introduction}
\label{sec:introduction}

Semantic communication (SemCom) shifts wireless design from exact bit
recovery toward task-relevant meaning and utility \cite{yang2023semcomsurvey}.
Deep learning enables end-to-end optimization of semantic representation,
physical-layer transmission, and task inference, as demonstrated by DeepSC
\cite{xie2021deepsc}. For images, DeepJSCC directly maps source samples to
channel symbols and degrades gracefully under bandwidth and channel mismatch
\cite{bourtsoulatze2019deepjscc}. Adaptive-rate and transformer-based
extensions further support variable bandwidth and SNR conditions
\cite{yang2022adaptive,bian2023deepjscclpp}.

Green SemCom, however, must account for computation as well as radio energy.
Existing work jointly optimizes semantic processing and wireless resources
\cite{yang2023energysemcom}, considers energy-efficient semantic edge
operation \cite{zheng2024aerialgreen}, and develops probabilistic green
SemCom with joint communication--computation optimization
\cite{xu2025greenpsc}. Under adversarial corruption, this trade-off becomes
more difficult because receiver-side defense can itself be energy- and
latency-intensive. Prior studies have established source- and physical-layer
semantic attacks and corresponding robust-training mechanisms
\cite{hu2023robustvqvae,nan2023physical,li2023semblk}; recent analysis also
shows that DeepJSCC can retain non-negligible intrinsic robustness under
noise-aware training \cite{zhang2026unanticipated}.

Diffusion-based secure SemCom provides adaptive recovery but does not by
itself solve the green cross-layer control problem. DiffuSeC adapts diffusion
depth to channel conditions \cite{ren2023diffusec}, while energy-aware
generative defense jointly considers transmission and defense energy in a
secure semantic setting \cite{zheng2024secureenergy}. ROME estimates semantic
jamming conditions and selects robust models \cite{zhou2026rome}, and SecDiff
uses adaptive diffusion guidance against physical-layer attacks
\cite{zhao2026secdiff}. Our recent TwinViT-DeepJSCC preprint
\cite{fallahreyhani2026twinvit} instead develops a preventive--corrective
transceiver combining twin ViT representations, sensitivity-aware masking,
confidence-aware fusion, blind corruption-severity estimation, and adaptive
DDIM purification under a fixed aggregate channel-use budget. That work
focuses on architectural robustness; it does not formulate long-term joint
minimization of radio and defensive-computation energy under partially
observable adversarial dynamics and semantic tail-risk constraints.

Adaptive resource control and risk-sensitive learning provide complementary
tools. Hybrid-action PPO has been used for semantic compression and resource
allocation \cite{lin2025rlsemcom}, while primal--dual PPO has been used for
constraint-aware semantic adaptation \cite{li2026latency}. CVaR-based resource
management has also appeared in secrecy-oriented semantic scheduling
\cite{lang2026meaningfirst}, and CVaR more generally provides a principled
mechanism for controlling low-probability, high-consequence outcomes
\cite{chow2018risk}. These ingredients motivate an energy-first formulation
in which average semantic quality, latency, and trajectory-level semantic risk
remain explicit constraints rather than fixed reward weights.

Table~\ref{tab:related_work_green} positions the proposed framework against
the closest green, secure, and adaptive SemCom approaches. The main gap is the
absence of a unified sequential formulation that simultaneously minimizes
communication and defensive-computation energy, adapts radio rate/power and
purification depth from causal observations, and enforces a trajectory-level
semantic CVaR constraint under hidden adversarial conditions.

\begin{table}[t]
\centering
\caption{Comparison With Closely Related Green, Secure, and Adaptive Semantic Communication Frameworks}
\label{tab:related_work_green}
\footnotesize
\renewcommand{\arraystretch}{0.98}
\setlength{\tabcolsep}{1.8pt}
\begin{tabularx}{\textwidth}{@{}*{7}{>{\raggedright\arraybackslash}X}@{}}
\toprule
\textbf{Method} & \textbf{Threat / Setting} & \textbf{Energy Treatment} &
\textbf{Adaptive Decisions} & \textbf{Adversarial-State Treatment} &
\textbf{Tail Risk} & \textbf{Main Difference From This Work}\\
\midrule
Yang \emph{et al.} \cite{yang2023energysemcom} & No explicit adversary &
Communication--computation energy minimization & Semantic extraction,
compute frequency, radio resources & Not applicable & No & Green SemCom
without adversarial defense\\
Xu \emph{et al.} \cite{xu2025greenpsc} & Multi-user probabilistic SemCom &
Total communication--computation energy minimization & Power, beamforming,
compression, compute capacity & No adversarial state & No & No active attacks
or defensive-computation control\\
Zheng \emph{et al.} \cite{zheng2024secureenergy} & Adversarial SIoT image
retrieval & Transmission and generative-defense energy & Energy-aware
transmission and defense & No hidden sequential attack state & No & No
partially observable cross-layer control or semantic CVaR\\
DiffuSeC \cite{ren2023diffusec} & Source/channel semantic attacks & No
long-term total-energy objective & Diffusion-timestep adaptation & No
probabilistic hidden attack belief & No & No joint radio--rate--defense energy
minimization\\
ROME \cite{zhou2026rome} & Power-variable semantic jamming & No joint
communication--defense energy objective & Robust-model ensembling & Attack
presence/power estimated & No & Attack-aware model selection rather than
green sequential control\\
SecDiff \cite{zhao2026secdiff} & Subcarrier jamming/pilot spoofing &
Computational-cost/latency efficiency & Adaptive diffusion guidance/sampling &
Attack-specific recovery & No & No long-term communication--defense energy
minimization\\
TwinViT-DeepJSCC \cite{fallahreyhani2026twinvit} & Source/channel active
attacks & Fixed channel-use budget; no long-term energy objective & Latent
masking, fusion, severity-adaptive DDIM & Blind severity estimate; no recurrent
hidden-state control & No & Preventive--corrective transceiver rather than
energy-minimizing sequential control\\
Lang and Ji \cite{lang2026meaningfirst} & Passive eavesdropper & Power and
edge-compute orchestration & Bandwidth, power, compute cycles & Partial
secrecy-state information & CVaR for secrecy & No active perturbations or
diffusion-defense energy\\
\textbf{Proposed} & \textbf{Source, OTA, and composite active attacks} &
\textbf{Communication + defensive-computation energy minimization} &
\textbf{Transmit power, semantic bandwidth, purification complexity} &
\textbf{Imperfect attack belief, uncertainty, recurrent history} &
\textbf{Semantic CVaR} & \textbf{Energy-minimizing risk-aware defense under
partial adversarial observability}\\
\bottomrule
\end{tabularx}
\end{table}

Motivated by this gap, we formulate adversarial SemCom as an
energy-minimizing constrained partially observable Markov decision process
(CPOMDP). The controller minimizes long-term communication and
purification energy while enforcing average semantic-loss, trajectory-level
$\mathrm{CVaR}_{\alpha}$, and end-to-end latency constraints. The true attack
mode and instantaneous attack strength are hidden; decisions rely on causal
channel information, semantic confidence, an imperfect attack-belief
distribution and its uncertainty, previous actions, and recurrent history.

The main contributions are:
\begin{itemize}
\item \textbf{Energy-aware adversarial SemCom model:} a sequential
DeepJSCC system with source, OTA, and composite attack modes that explicitly
accounts for radio and diffusion-purification energy.
\item \textbf{Energy-first risk-constrained formulation:} total energy is
the primary objective, while average semantic loss, trajectory-level CVaR,
and latency remain separate service constraints.
\item \textbf{Partially observable hybrid control:} a recurrent controller
jointly adapts continuous transmit power and discrete bandwidth/purification
actions from causal channel, confidence, attack-belief, uncertainty, and
history information.
\item \textbf{Recurrent primal--dual PPO and green robustness evaluation:}
a GRU-based hybrid actor with multi-head critics and adaptive dual variables
is paired with a CIFAR-100 evaluation protocol for source-domain and OTA
attacks, including held-out attack strengths. The experimental analysis evaluates resource accounting and measured performance; the system model additionally
supports composite source--OTA corruption.
\end{itemize}

The remainder of the paper presents the system model, the recurrent
primal--dual solution, the experimental methodology, measured performance analysis, and conclusion.


\section{System Model}
\label{sec:system_model}

We consider a point-to-point semantic image communication system
operating over a finite horizon of $T$ transmission intervals in the
presence of time-varying wireless channels and uncertain adversarial
perturbations. The legitimate system consists of a rate-adaptive
DeepJSCC transceiver, a wireless channel, a diffusion-based
receiver-side purifier, a frozen downstream semantic task network, and
a recurrent cross-layer controller. In contrast to conventional formulations that treat
receiver-side defense as computationally free, the proposed model
explicitly accounts for both the radio energy required for semantic
transmission and the computation energy consumed by iterative
diffusion purification.

\begin{figure}[!t]
    \centering
    \includegraphics[width=0.55\textwidth,height=0.65\textheight,keepaspectratio]{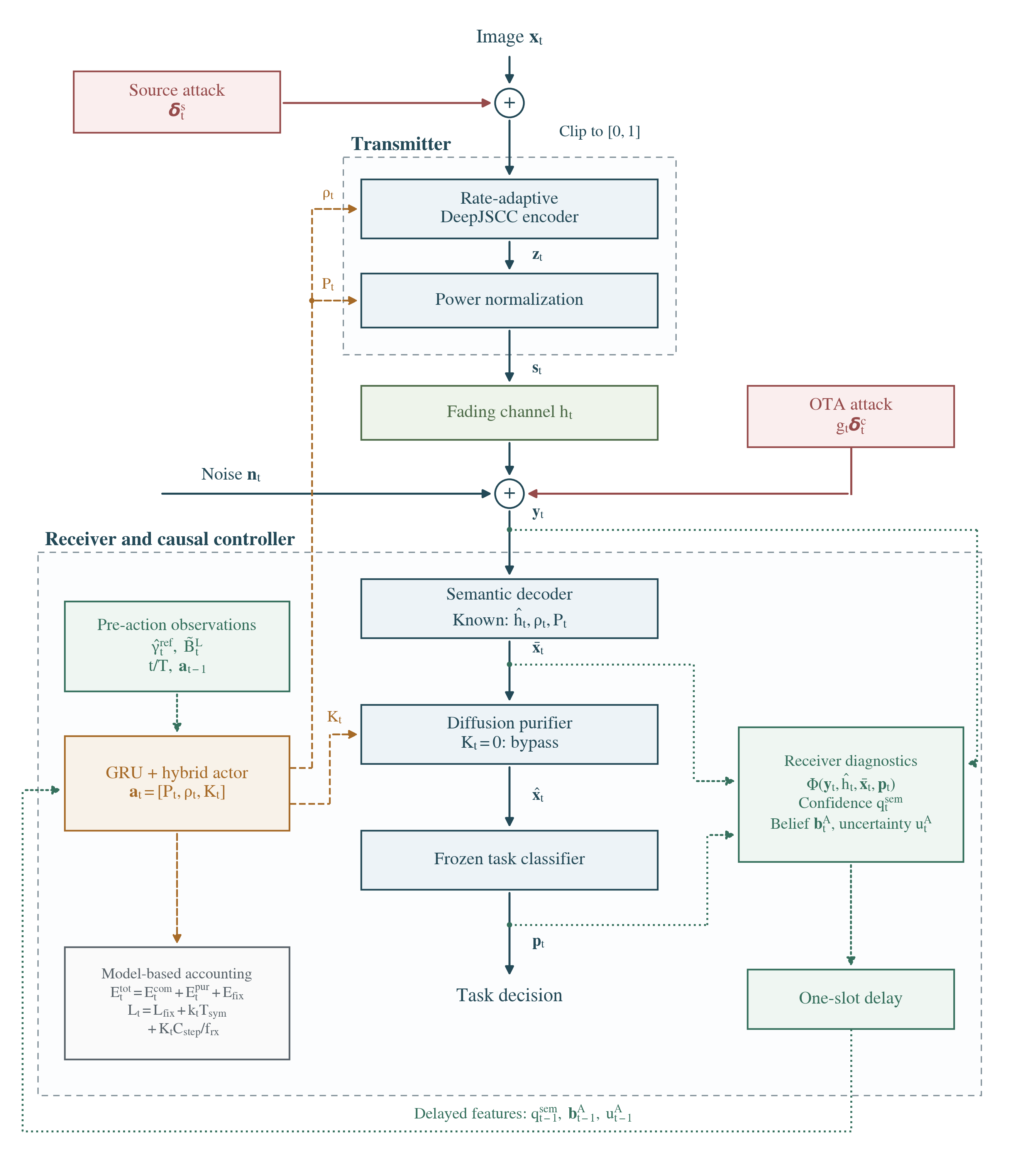}
    \caption{System model of the proposed energy-efficient,
    risk-constrained adaptive semantic communication framework.}
    \label{fig:system_model}
\end{figure}

The overall system architecture is illustrated in
Fig.~\ref{fig:system_model}, including the semantic transmission chain,
source-domain and OTA attacks, and the causal recurrent controller
that jointly selects $P_t$, $\rho_t$, and $K_t$.

At each transmission interval, the controller jointly determines the
physical-layer transmit power, the semantic/channel bandwidth ratio,
and the purification complexity. These decisions create a coupled
robustness--energy--latency trade-off: increasing transmit power or
bandwidth generally improves communication reliability but increases
radio energy consumption, whereas increasing the number of diffusion
steps can improve adversarial robustness at the expense of additional
receiver-side computation energy and processing delay.

The controller is restricted to causally available legitimate-system
information. In particular, the true adversarial mode, attack strength,
and realized perturbations are hidden during deployment. The receiver
therefore relies on channel-quality information, semantic confidence,
an imperfect attack-belief distribution, and recurrent observation
history to determine whether additional communication or defensive
energy expenditure is justified. A low-rate authenticated feedback link
is assumed for conveying channel-quality information and
transmitter-side control decisions. Its communication energy and delay
are assumed negligible compared with the semantic data transmission and
are not included in the optimization. The main notation used throughout
the formulation is summarized in Table~\ref{tab:notation}.


\begin{table}[t]
	\centering
	\caption{Main Notation}
	\label{tab:notation}
	\footnotesize
	\renewcommand{\arraystretch}{0.98}
	\setlength{\tabcolsep}{2.2pt}
	\begin{tabular}{@{}p{0.15\textwidth}p{0.32\textwidth}p{0.15\textwidth}p{0.32\textwidth}@{}}
		\toprule
		\textbf{Symbol} & \textbf{Definition} & \textbf{Symbol} & \textbf{Definition} \\
		\midrule
		$t,T$ & Transmission index and finite-horizon length & $\mathbf{x}_t,\mathbf{x}_t^{a}$ & Clean source image and encoder input after source perturbation \\
		$n=3HW$ & Number of real-valued source samples & $c_t,N_c$ & Ground-truth class and number of task classes \\
		$f_\theta,g_\phi$ & Semantic encoder and decoder & $\mathbf{z}_t,\mathbf{s}_t$ & Selected latent vector and transmitted complex signal \\
		$\rho_t,k_t,\mathcal R$ & Bandwidth ratio, symbol count, and ratio action set & $P_t,P_{\min},P_{\max}$ & Radiated transmit power and its bounds \\
		$K_t,\mathcal K$ & Reverse-diffusion step count and purification action set & $\mathcal D_\psi,q_\omega$ & Diffusion purifier and frozen semantic task network \\
		$h_t,g_t,\hat h_t$ & Legitimate channel, adversarial channel, and legitimate-channel estimate & $\mathbf y_t,\sigma_n^2$ & Received signal and noise variance per complex component \\
		$m_t^A,\mathcal M$ & Attack mode and attack-mode set & $\mathbf\Pi_A$ & Attack-mode transition matrix \\
		$\boldsymbol\delta_t^s,\epsilon_t^s$ & Source perturbation and its $\ell_\infty$ budget & $\boldsymbol\delta_t^c,P_t^A$ & OTA perturbation and its average transmit-power budget \\
		$\eta_h,\eta_g$ & Temporal correlation coefficients of legitimate and adversarial channels & $P_{\mathrm{ref}},\hat\gamma_t^{\mathrm{ref}}$ & Reference transmit power and corresponding channel-quality estimate \\
		$\bar{\mathbf x}_t,\hat{\mathbf x}_t$ & Preliminary and final reconstructed images & $\mathbf p_t,q_t^{\mathrm{sem}}$ & Class-posterior vector and entropy-based semantic confidence \\
		$\mathbf b_t^A,u_t^A$ & Estimated attack-mode distribution and its normalized entropy & $\mathbf o_t,\mathbf r_t$ & Causal observation and recurrent history representation \\
		$\mathcal H_t$ & Causal observation--action history before action selection & $\mathbf a_t$ & Hybrid action $[P_t,\rho_t,K_t]$ \\
		$\tilde B_t^L$ & Normalized remaining latency-budget slack & $T_{\mathrm{sym}}$ & Duration of one complex channel symbol \\
		$\eta_{\mathrm{PA}},P_{\mathrm c}$ & Power-amplifier efficiency and aggregate active radio-circuit power & $E_t^{\mathrm{com}},E_t^{\mathrm{pur}}$ & Communication and purification energy \\
		$E_{\mathrm{fix}},E_t^{\mathrm{tot}}$ & Fixed processing energy and total per-interval energy & $C_{\mathrm{step}},f_{\mathrm{rx}},\kappa_{\mathrm{rx}}$ & Cycles per diffusion step, processing frequency, and processing-energy coefficient \\
		$L_t^{\mathrm{pur}},L_t$ & Purification and total end-to-end latency & $L_{\mathrm{avg}}$ & Upper bound on expected average end-to-end latency \\
		$\ell_t^{\mathrm{rec}},\ell_t^{\mathrm{task}}$ & Reconstruction loss and task negative log-likelihood & $\ell_t^{\mathrm{sem}}$ & Combined reconstruction and task loss \\
		$\lambda_{\mathrm{rec}},\lambda_{\mathrm{task}}$ & Nonnegative reconstruction and task-loss weights & $J_E(\pi),$ $J_S(\pi),$ $J_L(\pi)$ & Expected average energy, semantic loss, and latency \\
		$\gamma,D_T$ & Semantic-loss discount factor and normalizer $D_T=\sum_{t=0}^{T-1}\gamma^t$ & $Z_\pi$ & Discount-weighted trajectory-average semantic loss \\
		$\alpha,\nu$ & CVaR confidence level and auxiliary VaR threshold & $\tau_{\mathrm{sem}},\tau_{\mathrm{risk}}$ & Average semantic-loss limit and trajectory-CVaR limit \\
		$\Omega,\Theta,\Psi$ & Shared-backbone, actor, and critic parameter collections & $E_{\mathrm{ref}},\ell_{\mathrm{ref}},L_{\mathrm{ref}}$ & Fixed energy, semantic-loss, and latency normalization references \\
		$c_t^E,c_t^S,c_t^L$ & Normalized instantaneous energy, semantic loss, and latency & $d_S,d_L$ & Normalized average semantic-loss and latency limits \\
		$V_\Psi^E,V_\Psi^S,V_\Psi^L,V_\Psi^R$ & Value heads for negative energy return, semantic cost, latency cost, and tail excess & $\widehat A_t^E,\widehat A_t^S,\widehat A_t^L,\widehat A_{i,t}^R$ & Energy, semantic, latency, and Monte Carlo risk advantages \\
		$\lambda_{\mathrm{GAE}}$ & GAE trace parameter for additive return streams & $N_{\mathrm{ep}},\widehat\nu_\alpha$ & Complete trajectories per batch and empirical VaR threshold \\
		$\zeta_i^R$ & Trajectory tail excess $[Z_i-\widehat\nu_\alpha]_+/(1-\alpha)$ & $\mu_S,\mu_L,\mu_R$ & Nonnegative semantic, latency, and CVaR multipliers \\
		$\varrho_t(\Theta)$ & New-to-old policy density ratio conditioned on the same causal history & $\varepsilon_{\mathrm{clip}}$ & PPO clipping parameter \\
		\bottomrule
	\end{tabular}
\end{table}


Additional notation used below is as follows: $D_T$ is the discount-weight
normalizer, $\mathcal H_t$ the causal pre-action history, and $t/T$ the
normalized time. The estimated attack probability is
$\widehat P_{A,t}=1-b_t^0$, distinct from attacker transmit power $P_t^A$.
$\Gamma_A$ is the attack/reference-power ratio in dB; $N_{\mathrm{test}}$
is the test-set size. $L_{\mathrm{fix}}$ denotes aggregate non-purification processing latency.

\subsection{Rate-Adaptive Semantic Transmission}
\label{subsec:semantic_transceiver}

At transmission interval $t$, the source image is denoted by
$\mathbf{x}_t\in[0,1]^{H\times W\times3}$, with $n=3HW$ source
samples. Depending on the active adversarial mode, the semantic encoder
processes either the clean source image or its adversarially perturbed
counterpart $\mathbf{x}_t^{a}$.

The controller selects the semantic/channel bandwidth ratio from the
finite set
\begin{equation}
	\rho_t
	\in
	\mathcal{R}
	=
	\{\rho_1,\ldots,\rho_M\},
	\qquad
	k_t
	=
	\left\lceil
	\rho_t n
	\right\rceil ,
	\label{eq:rate_selection}
\end{equation}
where every $\rho\in\mathcal R$ is positive and $k_t$ denotes the number of active complex channel symbols. The
selected ratio therefore affects both semantic representation capacity
and the duration, and consequently the energy cost, of the wireless
transmission.

The corresponding rate-adaptive latent representation is
\begin{equation}
	\mathbf{z}_t
	=
	\mathcal{S}_{\rho_t}
	\left(
	f_{\theta}
	\left(
	\mathbf{x}_t^{a}
	\right)
	\right)
	\in
	\mathbb{C}^{k_t},
	\label{eq:semantic_encoding}
\end{equation}
where $\mathcal{S}_{\rho_t}(\cdot)$ denotes the rate-selection
operation associated with $\rho_t$.

The controller simultaneously selects the radiated transmit power
$P_t\in[P_{\min},P_{\max}]$. The transmitted semantic vector is
normalized according to
\begin{equation}
	\mathbf{s}_t
	=
	\sqrt{k_tP_t}\,
	\frac{\mathbf{z}_t}
	{\|\mathbf{z}_t\|_2},
	\label{eq:transmit_signal}
\end{equation}
For $\|\mathbf{z}_t\|_2>0$, this guarantees
$\frac{1}{k_t} \|\mathbf{s}_t\|_2^2 = P_t.$
Thus, $P_t$ controls the average radiated power per transmitted complex
symbol without modifying the semantic encoder parameters.


\subsection{Composite Adversarial and Wireless Channel Model}
\label{subsec:threat_model}

We consider four possible adversarial operating modes:
$m_t^{A} \in \mathcal{M} = \{0,S,C,SC\},$
where $0$, $S$, $C$, and $SC$ represent benign operation,
source-domain attack, channel-domain attack, and simultaneous
source--channel attack, respectively.

For a source-domain attack, the semantic encoder input is modeled as
\begin{equation}
	\mathbf{x}_t^{a}
	=
	\Pi_{[0,1]}
	\left(
	\mathbf{x}_t
	+
	I_S(m_t^{A})
	\boldsymbol{\delta}_t^{s}
	\right),
	\qquad
	\|
	\boldsymbol{\delta}_t^{s}
	\|_{\infty}
	\leq
	\epsilon_t^{s},
	\label{eq:source_attack}
\end{equation}
where $\Pi_{[0,1]}(\cdot)$ projects its argument onto the valid image
range and
\[
I_S(m_t^{A})
	=
	\begin{cases}
		1, & m_t^{A}\in\{S,SC\},\\
		0, & \text{otherwise}.
	\end{cases}
\]

The received wireless signal is modeled as
\begin{equation}
	\mathbf{y}_t
	=
	h_t\mathbf{s}_t
	+
	I_C(m_t^{A})
	g_t\boldsymbol{\delta}_t^{c}
	+
	\mathbf{n}_t,
	\label{eq:received_signal}
\end{equation}
where
\[
I_C(m_t^{A})
	=
	\begin{cases}
		1, & m_t^{A}\in\{C,SC\},\\
		0, & \text{otherwise},
	\end{cases}
\]
and
$\mathbf{n}_t \sim \mathcal{CN} \left( \mathbf{0}, \sigma_n^2\mathbf{I}_{k_t} \right).$
The OTA adversarial waveform
$\boldsymbol{\delta}_t^{c}\in\mathbb{C}^{k_t}$ satisfies
\begin{equation}
	\frac{1}{k_t}
	\|
	\boldsymbol{\delta}_t^{c}
	\|_2^2
	\leq
	P_t^{A}.
	\label{eq:ota_attack_constraint}
\end{equation}

\subsubsection{Adversary Knowledge and Attack Objective}
\label{subsubsec:adversary_objective}

The adversary is modeled as an end-to-end semantic attacker whose
objective is to degrade the semantic utility of the legitimate
communication link rather than merely increase received-signal
distortion. The source-domain and OTA perturbations are therefore
constructed to increase the downstream semantic task loss subject to
their corresponding perturbation or power constraints.

The source-domain adversarial perturbation is represented generically
as
\begin{equation}
	\boldsymbol{\delta}_{t}^{s,\star}
	\in
	\arg\max_{
		\|
		\boldsymbol{\delta}
		\|_{\infty}
		\leq
		\epsilon_t^{s}}
	\mathcal{L}_{\mathrm{adv}}^{s}
	\left(
	\mathbf{x}_t,
	\boldsymbol{\delta};
	\mathcal{F}
	\right),
	\label{eq:source_attack_objective}
\end{equation}
where $\mathcal{F}$ denotes the complete semantic communication and
inference pipeline.

Similarly, the OTA attack is represented by
\begin{equation}
	\begin{aligned}
		\boldsymbol{\delta}_{t}^{c,\star}
		\in {}&
		\arg\max_{\boldsymbol{\delta}}
		\;
		\mathcal{L}_{\mathrm{adv}}^{c}
		\left(
		\mathbf{x}_t,
		\boldsymbol{\delta};
		\mathcal{F},
		h_t,
		g_t
		\right)
		\\
		\mathrm{s.t.}\quad&
		\frac{1}{k_t}
		\|
		\boldsymbol{\delta}
		\|_2^2
		\leq
		P_t^{A}.
	\end{aligned}
	\label{eq:ota_attack_objective}
\end{equation}

In the composite mode $SC$, both perturbations may be simultaneously
present and independently satisfy
\eqref{eq:source_attack} and
\eqref{eq:ota_attack_constraint}.

Importantly, the legitimate controller does not directly observe
$m_t^{A}$, $\epsilon_t^{s}$, $P_t^{A}$, or the realized adversarial
perturbations. These variables belong to the hidden environment state.
The controller must instead infer the underlying operating condition
from causal receiver-side observations and the imperfect attack-belief
information introduced in Section~\ref{subsec:belief_estimation}.


The attack objective is the downstream semantic negative log-likelihood,
$\mathcal L_{\mathrm{adv}}^{d}=\mathbb E[-\log p_{t,c_t}^{(d)}]$,
$d\in\{s,c\}$. During attack generation, the adversary is white-box with
respect to the fixed semantic transceiver, downstream classifier, current
channel realization, and selected action, but it has no access to the
controller recurrent state, future actions, or future channel/attack states.
Source and OTA perturbations are projected onto their respective constraints
after every gradient step. When the composite mode is instantiated, source
and OTA perturbations are updated alternately under their separate budgets.
None of this ground-truth attack information is revealed to the deployed
controller.

\subsection{Temporal Channel and Attack Dynamics}
\label{subsec:environment_dynamics}

To capture temporal correlation in wireless propagation, the legitimate
channel coefficient follows a first-order Gauss--Markov process:
\begin{equation}
	h_{t+1}
	=
	\eta_h h_t
	+
	\sqrt{1-\eta_h^2}\,e_t,
	\qquad
	e_t
	\sim
	\mathcal{CN}(0,1),
	\label{eq:gauss_markov_channel}
\end{equation}
where $0\leq\eta_h<1$ determines the legitimate-channel correlation.

Similarly, the adversary-to-receiver channel evolves as
\begin{equation}
	g_{t+1}
	=
	\eta_g g_t
	+
	\sqrt{1-\eta_g^2}\,v_t,
	\qquad
	v_t
	\sim
	\mathcal{CN}(0,1),
	\label{eq:gauss_markov_adversary}
\end{equation}
where $0\leq\eta_g<1$ is the adversarial-channel correlation
coefficient. We initialize
$h_0\sim\mathcal{CN}(0,1)$ and
$g_0\sim\mathcal{CN}(0,1)$ so that both processes have unit stationary
variance.

The adversarial operating mode evolves according to a first-order
Markov process:
\begin{equation}
	\Pr
	\left(
	m_{t+1}^{A}=j
	\mid
	m_t^{A}=i
	\right)
	=
	[\mathbf{\Pi}_A]_{ij},
	\qquad
	i,j\in\mathcal{M},
	\label{eq:attack_markov}
\end{equation}
where $\mathbf{\Pi}_A$ is row stochastic:
$[\mathbf{\Pi}_A]_{ij} \geq 0, \qquad \sum_{j\in\mathcal{M}} [\mathbf{\Pi}_A]_{ij} = 1.$

During training, source images are sampled uniformly from the CIFAR-100
training split. Attack modes follow the Markov chain in
\eqref{eq:attack_markov}, and the active attack strength is sampled from the
training-strength set reported in Section~\ref{subsec:experimental_attacks}.
Both quantities remain hidden from the controller. Temporal correlation in
wireless propagation and adversarial activity motivates the recurrent policy.


\subsection{Semantic Reconstruction and Adaptive Purification}
\label{subsec:purification}

The receiver first forms a preliminary semantic reconstruction as
\begin{equation}
\bar{\mathbf{x}}_t=g_{\phi}(\mathbf{y}_t,\hat h_t,\rho_t,P_t).
\label{eq:semantic_decoder}
\end{equation}
Here, $\hat h_t$ is the legitimate-channel estimate; the selected
$P_t$ and $\rho_t$ are known at the receiver through authenticated control
signaling.

The controller selects the purification complexity from
$K_t \in \mathcal{K} = \{0,K_1,\ldots,K_Q\},$
where $K_t=0$ bypasses the diffusion purifier. The final reconstructed
image is
\begin{equation}
	\hat{\mathbf{x}}_t
	=
	\mathcal{D}_{\psi}
	\left(
	\bar{\mathbf{x}}_t;
	K_t
	\right).
	\label{eq:purification}
\end{equation}

The frozen downstream semantic task network then produces
\begin{equation}
	\mathbf{p}_t
	=
	q_{\omega}
	\left(
	\hat{\mathbf{x}}_t
	\right)
	\in
	[0,1]^{N_c},
	\qquad
	\sum_{c=1}^{N_c}
	p_{t,c}
	=
	1.
	\label{eq:task_posterior}
\end{equation}

The selection of $K_t$ therefore affects not only semantic robustness
but also the computational energy and latency of receiver-side defense,
as modeled next.


\subsection{Communication and Defensive-Computation Energy Model}
\label{subsec:energy_model}

A central distinction of the proposed formulation is that energy is
treated as the primary optimization objective rather than as an
auxiliary resource constraint. We explicitly separate radio
communication energy from the computation energy consumed by
diffusion-based semantic defense.

For an active transmission of $k_t$ complex symbols, the communication
energy is modeled as
\begin{equation}
	E_t^{\mathrm{com}}
	=
	\left(
	\frac{P_t}{\eta_{\mathrm{PA}}}
	+
	P_{\mathrm{c}}
	\right)
	k_t T_{\mathrm{sym}},
	\label{eq:communication_energy}
\end{equation}
where $\eta_{\mathrm{PA}}\in(0,1]$ is the power-amplifier efficiency and
$P_{\mathrm{c}}$ represents the aggregate transmitter/receiver circuit
power while the radio link is active. Unlike the radiated-energy model
$P_tk_tT_{\mathrm{sym}}$, \eqref{eq:communication_energy} accounts for
both power-amplifier inefficiency and active radio circuitry.

To model defensive computation, let $C_{\mathrm{step}}$ denote the
equivalent number of processing cycles required for one reverse
diffusion step. The total number of purification cycles is
$C_t^{\mathrm{pur}} = K_t C_{\mathrm{step}}.$

Using the standard dynamic-processing energy model adopted in
energy-efficient semantic and edge-computing systems
\cite{yang2023energysemcom}, the purification energy is
\begin{equation}
	E_t^{\mathrm{pur}}
	=
	\kappa_{\mathrm{rx}}
	f_{\mathrm{rx}}^{2}
	C_t^{\mathrm{pur}}
	=
	\kappa_{\mathrm{rx}}
	f_{\mathrm{rx}}^{2}
	K_t C_{\mathrm{step}},
	\label{eq:purification_energy}
\end{equation}
where $f_{\mathrm{rx}}$ denotes the effective receiver processing
frequency and $\kappa_{\mathrm{rx}}$ is the effective processing-energy model
coefficient. The coefficient combination defines a processor-level surrogate, not
measured GPU energy. Mapping it to an implemented neural purifier requires
profiling; the present numerical coefficients are declared modeling
assumptions and are not a calibration of the RTX~4070.

The effective frequency $f_{\mathrm{rx}}$ is treated as a fixed hardware
parameter in this work rather than an additional control action. This
choice preserves the original three-dimensional hybrid action space and
isolates the principal trade-off between radio-resource allocation and
adaptive defensive computation.

The corresponding purification latency is
\begin{equation}
	L_t^{\mathrm{pur}}
	=
	\frac{C_t^{\mathrm{pur}}}
	{f_{\mathrm{rx}}}
	=
	\frac{
		K_t C_{\mathrm{step}}
	}
	{f_{\mathrm{rx}}},
	\label{eq:purification_latency}
\end{equation}
with
$E_t^{\mathrm{pur}} = L_t^{\mathrm{pur}} = 0, \qquad K_t=0.$

We use an action-independent aggregate processing-energy approximation
$E_{\mathrm{fix}}$ for the encoder, decoder, task network, attack estimator,
and controller. For the numerical accounting comparison, the same allowance
is assigned to every method. This is a controlled modeling assumption,
not evidence that their actual processing costs are identical. Training
energy, attacker energy, and external infrastructure are outside this
per-image inference boundary. Method- or rate-dependent measured overhead
would require $E_{\mathrm{fix}}^{(m)}(\rho)$ and corresponding latency
terms to be used consistently in both optimization and evaluation.

The total per-interval energy is therefore
\begin{equation}
	E_t^{\mathrm{tot}}
	=
	E_t^{\mathrm{com}}
	+
	E_t^{\mathrm{pur}}
	+
	E_{\mathrm{fix}}.
	\label{eq:total_energy}
\end{equation}

The total end-to-end latency is modeled as
\begin{equation}
\begin{aligned}
L_t&=L_{\mathrm{fix}}+k_tT_{\mathrm{sym}}+K_tC_{\mathrm{step}}/f_{\mathrm{rx}},\\
L_{\mathrm{fix}}&=L_t^{\mathrm{enc}}+L_t^{\mathrm{dec}}+L_t^{\mathrm{task}}+L_t^{\mathrm{ctrl}}.
\end{aligned}
\label{eq:end_to_end_latency}
\end{equation}

The constant $L_{\mathrm{fix}}$ includes attack estimation and control
signaling processing in the resource model; queueing and retransmission
are excluded. Energy and latency must be averaged over the same actions.
Let $k_{\min}=\lceil n\rho_{\min}\rceil$ and
$k_{\max}=\lceil n\rho_{\max}\rceil$. The model implies
\begin{equation}
\begin{aligned}
E_{\min}&=E_{\mathrm{fix}}+(P_{\min}/\eta_{\mathrm{PA}}+P_{\mathrm c})k_{\min}T_{\mathrm{sym}},\\
E_{\max}&=E_{\mathrm{fix}}+(P_{\max}/\eta_{\mathrm{PA}}+P_{\mathrm c})k_{\max}T_{\mathrm{sym}}
+\kappa_{\mathrm{rx}}f_{\mathrm{rx}}^2K_{\max}C_{\mathrm{step}}.
\end{aligned}
\label{eq:energy_bounds}
\end{equation}
For $K=0$, the last term is absent. These bounds also hold for sample
averages and for normalized energy after division by the common
$E_{\mathrm{ref}}$. Under the fixed-latency approximation, average latency
feasibility additionally requires
\begin{equation}
\mathbb E[K]\leq\frac{f_{\mathrm{rx}}}{C_{\mathrm{step}}}
\left(L_{\mathrm{avg}}-L_{\mathrm{fix}}-T_{\mathrm{sym}}\mathbb E[k]\right).
\label{eq:mean_k_bound}
\end{equation}
This is an average-action condition, not a prohibition on occasional
$K=20$ actions and not a guarantee of semantic feasibility.

Equations~\eqref{eq:communication_energy}--\eqref{eq:end_to_end_latency}
reveal the coupled role of the hybrid control variables. Increasing
$P_t$ directly increases radio energy, increasing $\rho_t$ increases
the number of transmitted symbols and therefore both transmission
duration and communication energy, while increasing $K_t$ raises both
defensive-computation energy and purification latency.


\subsection{Causal Semantic Confidence and Attack-Belief Estimation}
\label{subsec:belief_estimation}

The deployed controller is restricted to information that can be
obtained causally by the legitimate system. In particular, the clean
source image, ground-truth semantic label, true adversarial mode,
instantaneous attack strength, and ground-truth reconstruction error
are excluded from the controller observation.

The semantic-confidence score is computed from the semantic-task
posterior as
\begin{equation}
	q_t^{\mathrm{sem}}
	=
	1
	+
	\frac{1}{\log N_c}
	\sum_{c=1}^{N_c}
	p_{t,c}
	\log p_{t,c},
	\qquad
	0
	\leq
	q_t^{\mathrm{sem}}
	\leq
	1,
	\label{eq:semantic_confidence}
\end{equation}
where larger values indicate lower posterior entropy and therefore
higher confidence in the reconstructed semantic content. We use the
standard convention $0\log0=0$.

A lightweight receiver-side estimator processes only observable
features and produces the attack-belief distribution
\begin{equation}
	\mathbf{b}_t^{A}
	=
	d_{\chi}
	\left[
	\Phi
	\left(
	\mathbf{y}_t,
	\hat h_t,
	\bar{\mathbf{x}}_t,
	\mathbf{p}_t
	\right)
	\right]
	=
	[
	b_t^{0},
	b_t^{S},
	b_t^{C},
	b_t^{SC}
	],
	\label{eq:attack_belief}
\end{equation}
with
$b_t^{m} \geq 0, \qquad \sum_{m\in\mathcal{M}} b_t^{m} = 1.$

Here, $\Phi(\cdot)$ is a deterministic feature-extraction mapping and
$d_{\chi}(\cdot)$ is a lightweight attack-classification network with
softmax output.

The uncertainty associated with the inferred adversarial state is
measured through normalized entropy:
\begin{equation}
	u_t^{A}
	=
	-
	\frac{
		\displaystyle
		\sum_{m\in\mathcal{M}}
		b_t^{m}
		\log b_t^{m}
	}
	{\log|\mathcal{M}|},
	\qquad
	0
	\leq
	u_t^{A}
	\leq
	1.
	\label{eq:belief_uncertainty}
\end{equation}


\subsection{Partially Observable Cross-Layer Control}
\label{subsec:cpomdp}

For formulation purposes, the hidden environment state can be represented
conceptually as
\[\boldsymbol{\zeta}_t = \left[ \mathbf{x}_t, c_t, h_t, g_t, m_t^{A}, \epsilon_t^{s}, P_t^{A}, \tilde B_t^{L} \right].\]
This state is used only to characterize the environment and is not
directly available to the deployed controller. In particular, the
source and adversarial variables remain hidden.

Before action selection, the controller has access to the
reference-power channel-quality estimate
\[
\hat{\gamma}_t^{\mathrm{ref}}
	=
	\frac{
		|\hat h_t|^2
		P_{\mathrm{ref}}
	}
	{\sigma_n^2},
\]
where $P_{\mathrm{ref}}$ is a fixed reference transmit power. This
quantity provides a pre-decision indication of channel quality without
depending on the yet-unknown action $P_t$.

Because energy is now the optimization objective rather than a hard
budget, no remaining-energy variable is supplied to the controller.
Expected-average latency consumption is summarized by the normalized
remaining latency slack
\begin{equation}
	\tilde B_t^{L}
	=
	1
	-
	\frac{
		\displaystyle
		\sum_{\tau=0}^{t-1}
		L_{\tau}
	}
	{T L_{\mathrm{avg}}}.
	\label{eq:latency_slack}
\end{equation}

The causal observation available before selecting the action at
interval $t$ is
\begin{equation}
	\mathbf{o}_t
	=
	\left[
	\hat{\gamma}_t^{\mathrm{ref}},
	q_{t-1}^{\mathrm{sem}},
	\mathbf{b}_{t-1}^{A},
	u_{t-1}^{A},
	\tilde B_t^{L},
	t/T,
	\mathbf{a}_{t-1}
	\right].
	\label{eq:observation}
\end{equation}

The use of $q_{t-1}^{\mathrm{sem}}$, $\mathbf{b}_{t-1}^{A}$, and
$u_{t-1}^{A}$ prevents a causality violation because their current-slot
values become available only after transmission, reconstruction,
purification, and semantic inference have been completed. Similarly,
$\hat{\gamma}_t^{\mathrm{ref}}$ is used instead of the actual
post-action SNR, which depends on $P_t$.

The feature $t/T$ identifies the remaining finite horizon; the latency
slack is an observation feature rather than a pathwise delay guarantee. At
$t=0$, previous-slot components are initialized to neutral values.

Rather than constructing an exact Bayesian belief over the complete
hidden state, the controller maintains the recurrent representation
$\mathbf{r}_t = F_{\eta} \left( \mathbf{r}_{t-1}, \mathbf{o}_t \right),$
which summarizes the causal observation history.

The joint cross-layer action is
\begin{equation}
\mathbf a_t=[P_t,\rho_t,K_t],\quad P_t\in[P_{\min},P_{\max}],\quad\rho_t\in\mathcal R,\quad K_t\in\mathcal K.
\label{eq:joint_action}
\end{equation}

Hence, the action space remains hybrid: transmit power is continuous,
whereas semantic/channel bandwidth ratio and purification complexity
are discrete.


\subsection{Semantic Loss and Tail-Risk Model}
\label{subsec:semantic_risk}

During controller training, the simulator has access to the clean source
image $\mathbf{x}_t$ and its ground-truth semantic class $c_t$. These
quantities are used exclusively to construct the training losses and
constraints and are never included in the deployed controller
observation.

The reconstruction loss is
\begin{equation}
	\ell_t^{\mathrm{rec}}
	=
	\frac{
		\|
		\mathbf{x}_t
		-
		\hat{\mathbf{x}}_t
		\|_2^2
	}
	{3HW},
	\label{eq:reconstruction_loss}
\end{equation}
while the task-level semantic loss is
\begin{equation}
	\ell_t^{\mathrm{task}}
	=
	-\log
	p_{t,c_t}.
	\label{eq:task_loss}
\end{equation}

The combined semantic loss is
\begin{equation}
	\ell_t^{\mathrm{sem}}
	=
	\lambda_{\mathrm{rec}}
	\ell_t^{\mathrm{rec}}
	+
	\lambda_{\mathrm{task}}
	\ell_t^{\mathrm{task}},
	\label{eq:semantic_loss}
\end{equation}
where
$\lambda_{\mathrm{rec}}\geq0$ and
$\lambda_{\mathrm{task}}\geq0$
control the relative importance of reconstruction fidelity and
task-level semantic correctness, with
$\lambda_{\mathrm{rec}}+\lambda_{\mathrm{task}}>0$.
Task loss is evaluated using numerically stable log-softmax logits;
any probability clipping must also be reflected in this definition.

The expected average semantic loss under policy $\pi$ is
\begin{equation}
	J_S(\pi)
	=
	\mathbb{E}_{\pi}
	\left[
	\frac{1}{T}
	\sum_{t=0}^{T-1}
	\ell_t^{\mathrm{sem}}
	\right].
	\label{eq:average_semantic_loss}
\end{equation}

Average semantic performance alone may conceal a small number of severe
trajectory-level failures. We therefore define the discount-weighted trajectory-average semantic loss
\begin{equation}
Z_{\pi}=\frac{1}{D_T}\sum_{t=0}^{T-1}\gamma^t\ell_t^{\mathrm{sem}},
\qquad D_T=\sum_{t=0}^{T-1}\gamma^t,\quad 0<\gamma\leq1.
\label{eq:cumulative_semantic_loss}
\end{equation}
where $\gamma$ is the semantic-loss discount factor. Normalization by
$D_T$ makes the risk variable a weighted average on the same loss scale
as a single interval; for $\gamma=1$, it is the trajectory mean.
The threshold $\tau_{\mathrm{risk}}$ is defined on this normalized loss scale.

The corresponding conditional value-at-risk is
\begin{equation}
	\begin{aligned}
		\operatorname{CVaR}_{\alpha}
		\left(
		Z_{\pi}
		\right)
		=
		\min_{\nu\in\mathbb{R}}
		\left\{
		\nu
		+
		\frac{1}{1-\alpha}
		\mathbb{E}_{\pi}
		\left[
		\left(
		Z_{\pi}
		-
		\nu
		\right)_{+}
		\right]
		\right\},
		\\[
-1mm]
		0<\alpha<1,
	\end{aligned}
	\label{eq:cvar}
\end{equation}
where $(x)_{+}=\max(x,0)$ and $\nu$ is the auxiliary value-at-risk
parameter.
\subsection{Energy-Minimizing Risk-Constrained Optimization}
\label{subsec:energy_optimization}
The primary objective of the proposed green semantic communication
framework is to minimize the long-term energy expenditure required to
maintain reliable semantic service under uncertain adversarial
conditions. For policy $\pi$, the expected average total energy is
\begin{equation}
	J_E(\pi)
	=
	\mathbb{E}_{\pi}
	\left[
	\frac{1}{T}
	\sum_{t=0}^{T-1}
	E_t^{\mathrm{tot}}
	\right].
	\label{eq:average_total_energy}
\end{equation}
The corresponding expected average end-to-end latency is
\begin{equation}
	J_L(\pi)
	=
	\mathbb{E}_{\pi}
	\left[
	\frac{1}{T}
	\sum_{t=0}^{T-1}
	L_t
	\right].
	\label{eq:average_latency}
\end{equation}
The proposed adaptive semantic-defense problem is formulated as
\begin{equation}
	\boxed{
		\begin{aligned}
			\underset{\pi}{\operatorname{minimize}}
			\quad&
			J_E(\pi)
			\\[1mm]
			\mathrm{subject~to}
			\quad&
			J_S(\pi)
			\leq
			\tau_{\mathrm{sem}},
			\\
			&
			\operatorname{CVaR}_{\alpha}
			\left(
			Z_{\pi}
			\right)
			\leq
			\tau_{\mathrm{risk}},
			\\
			&
			J_L(\pi)
			\leq
			L_{\mathrm{avg}}.
	\end{aligned}}
	\label{eq:constrained_problem}
\end{equation}
The first constraint prevents energy minimization from causing
systematic degradation of semantic quality, while the second explicitly
limits the high-loss tail of the trajectory distribution. The latency
constraint bounds expected average delay; it does not guarantee a
per-packet deadline or a pathwise latency budget.
Consequently, the controller must learn when additional transmit energy,
semantic bandwidth, or diffusion-based defensive computation is
justified by the available evidence of poor channel quality or
adversarial activity. The resulting problem is an energy-minimizing,
risk-constrained hybrid CPOMDP whose solution is developed in
Section~\ref{sec:proposed_method}.
\section{Proposed Recurrent Energy-Minimizing Risk-Constrained Policy Optimization}
\label{sec:proposed_method}
The energy-minimization problem in
\eqref{eq:constrained_problem} poses four coupled learning challenges.
First, the legitimate controller does not directly observe the
instantaneous adversarial mode or attack strength and must instead
operate using imperfect causal evidence. Second, the cross-layer action
space is hybrid, consisting of a continuous transmit-power decision and
two discrete decisions for semantic bandwidth and purification
complexity. Third, minimizing energy alone may encourage overly
aggressive resource reduction and therefore must be performed while
enforcing average semantic-quality and latency requirements. Fourth,
the semantic-risk constraint is trajectory dependent because
$\operatorname{CVaR}_{\alpha}(Z_{\pi})$ depends on the tail of the
complete trajectory-loss distribution rather than on an additive
per-slot cost.
To address these challenges, we develop a recurrent primal--dual
proximal policy optimization (PPO) framework
\cite{schulman2017ppo}. A gated recurrent unit (GRU) extracts temporal
information from the partially observable interaction history, while a
hybrid stochastic actor jointly adapts transmit power, semantic/channel
bandwidth ratio, and diffusion-purification complexity. Unlike the
conventional reward formulation in which semantic utility is directly
maximized, the proposed actor is driven by an energy-minimization
reward. Semantic quality, end-to-end latency, and semantic tail risk
are enforced as separate constraints through adaptive dual variables.
A multi-head recurrent critic estimates the objective and constraint
returns required by the primal--dual update.
\subsection{Recurrent Representation Under Partial Observability}
\label{subsec:recurrent_representation}
The instantaneous observation $\mathbf{o}_t$ in
\eqref{eq:observation} does not uniquely determine the hidden
environment state. For example, a reduction in semantic confidence may
be caused by channel degradation, adversarial corruption, or both.
Similarly, a high attack-belief uncertainty may persist over multiple
intervals before sufficient evidence is accumulated to justify
energy-intensive defensive processing. Consequently, the current
observation alone is generally insufficient for energy-aware control.
We instantiate the recurrent mapping in
\eqref{eq:gru_update} using a GRU:
\begin{equation}
	\mathbf{r}_t
	=
	\operatorname{GRU}_{\eta}
	\left(
	\mathbf{o}_t,
	\mathbf{r}_{t-1}
	\right),
	\label{eq:gru_update}
\end{equation}
where $\eta$ denotes the GRU parameters. The recurrent state
$\mathbf{r}_t$ is not interpreted as an exact Bayesian posterior over
the hidden environment state. Instead, it provides a learned summary of
the causal observation history and captures temporal dependencies in
channel quality, semantic confidence, adversarial activity, attack
belief, belief uncertainty, previous control actions, and latency
consumption.
A shared feature representation is subsequently generated as
$\mathbf{f}_t = \varphi_{\eta_f} \left( \mathbf{r}_t \right),$
where $\eta_f$ denotes the parameters of the shared feature extractor.
The parameter collections satisfy $\Theta=(\Omega,\theta_{\mathrm{actor}})$
and $\Psi=(\Omega,\psi_{\mathrm{critic}})$; old collections contain frozen
copies of the backbone as well as their heads.
The recurrent encoder and shared feature extractor form the common
actor--critic backbone, denoted by $\Omega$. The shared parameters
receive gradients from both actor and critic objectives, whereas the
actor-specific and critic-specific output heads remain separate.
\subsection{Hybrid Energy-Aware Cross-Layer Actor}
\label{subsec:hybrid_actor}
The hybrid action
$\mathbf{a}_t = [P_t,\rho_t,K_t]$
contains one continuous component and two discrete components. Rather
than discretizing the entire joint action space, the actor preserves
this heterogeneous structure using one bounded continuous policy head
and two categorical policy heads.
Each action affects energy consumption through a different physical
mechanism. The transmit-power action $P_t$ controls power-amplifier
energy, $\rho_t$ determines the number of transmitted semantic symbols
and hence communication duration and circuit energy, while $K_t$
determines the amount of diffusion-based defensive computation.
Accordingly, the shared recurrent representation allows the agent to
coordinate communication and computation energy expenditures rather
than optimizing the three actions independently.
\subsubsection{Continuous Transmit-Power Policy}
To guarantee that the transmit-power action always remains within its
feasible interval, we employ a bounded Beta policy. A normalized power
variable is sampled according to
\begin{equation}
	u_t^{P}
	\sim
	\operatorname{Beta}
	\left(
	\kappa_{1,t}^{P},
	\kappa_{2,t}^{P}
	\right),
	\qquad
	0<u_t^{P}<1.
	\label{eq:beta_power}
\end{equation}
The positive Beta shape parameters are generated from the shared
feature representation:
\[
\begin{bmatrix}
		\kappa_{1,t}^{P}\\
		\kappa_{2,t}^{P}
	\end{bmatrix}
	=
	\operatorname{softplus}
	\left(
	\mathbf{W}_{P}\mathbf{f}_t
	+
	\mathbf{b}_{P}
	\right)
	+
	\varepsilon_{\mathrm{num}}\mathbf{1},
\]
where $\varepsilon_{\mathrm{num}}>0$ is a small constant for numerical
stability.

The normalized sample is mapped to the physical transmit-power range as
\begin{equation}
	P_t
	=
	P_{\min}
	+
	u_t^{P}
	\left(
	P_{\max}-P_{\min}
	\right).
	\label{eq:power_mapping}
\end{equation}
This construction satisfies the transmit-power bounds by design and
avoids post-sampling action clipping.

Assume $0\leq P_{\min}<P_{\max}$. The transformed density is
$\pi_P(P_t\mid\mathbf r_t)=f_{\mathrm{Beta}}(u_t^P;
\kappa_{1,t}^P,\kappa_{2,t}^P)/(P_{\max}-P_{\min})$.
The constant Jacobian cancels in the PPO ratio but is retained in
physical-action log densities.

\subsubsection{Discrete Bandwidth and Purification Policies}

The semantic/channel bandwidth ratio and purification complexity are
generated through independent categorical heads conditioned on the same
shared recurrent feature:
\begin{equation}
\begin{aligned}
	\boldsymbol{\pi}_{\rho,t}
	&=
	\operatorname{softmax}
	\left(
	\mathbf{W}_{\rho}\mathbf{f}_t
	+
	\mathbf{b}_{\rho}
	\right),
	&
	\rho_t
	&\sim
	\operatorname{Cat}
	\left(
	\boldsymbol{\pi}_{\rho,t}
	\right),
	\nonumber\\
	\boldsymbol{\pi}_{K,t}
	&=
	\operatorname{softmax}
	\left(
	\mathbf{W}_{K}\mathbf{f}_t
	+
	\mathbf{b}_{K}
	\right),
	&
	K_t
	&\sim
	\operatorname{Cat}
	\left(
	\boldsymbol{\pi}_{K,t}
	\right).
	\label{eq:discrete_actor_heads}
\end{aligned}
\end{equation}

Conditioned on $\mathbf{r}_t$, the joint stochastic policy is
factorized as
\begin{equation}
\pi_\Theta(\mathbf a_t\mid\mathbf r_t)=\pi_P(P_t\mid\mathbf r_t)\,\pi_\rho(\rho_t\mid\mathbf r_t)\,\pi_K(K_t\mid\mathbf r_t),
\label{eq:joint_policy}
\end{equation}
where $\Theta$ denotes the complete actor parameter collection.

Although the conditional action distribution is factorized for
computational tractability, all action components depend on the same
recurrent state and are optimized through a common energy-aware
primal--dual objective. The proposed controller therefore constitutes
a single hybrid-action agent rather than three independent policies.


\subsection{Energy Objective and Normalized Constraint Costs}
\label{subsec:reward_costs}

The optimization in \eqref{eq:constrained_problem} minimizes physical
total energy rather than semantic loss. For stable policy learning, we
introduce fixed positive normalization references
$E_{\mathrm{ref}}$, $\ell_{\mathrm{ref}}$, and $L_{\mathrm{ref}}$,
selected before controller training and kept unchanged across all
channel and attack conditions.

The normalized instantaneous quantities are defined as
\begin{equation}
	\begin{aligned}
		c_t^{E}
		&=
		\frac{
			E_t^{\mathrm{tot}}
		}
		{E_{\mathrm{ref}}},
		&
		c_t^{S}
		&=
		\frac{
			\ell_t^{\mathrm{sem}}
		}
		{\ell_{\mathrm{ref}}},
		&
		c_t^{L}
		&=
		\frac{
			L_t
		}
		{L_{\mathrm{ref}}}.
	\end{aligned}
	\label{eq:normalized_quantities}
\end{equation}

Because PPO is expressed as a reward-maximization algorithm, the
energy-minimization objective is represented by
$r_t^{E} = - c_t^{E} = - \frac{ E_t^{\mathrm{tot}} } {E_{\mathrm{ref}}}.$

The normalized limits associated with the average semantic and latency
constraints are
$d_S = \frac{ \tau_{\mathrm{sem}} } {\ell_{\mathrm{ref}}}$, $d_L = \frac{ L_{\mathrm{avg}} } {L_{\mathrm{ref}}}$.

Since all normalization factors are fixed positive constants, the
normalized formulation preserves the optimizer and feasible set of
\eqref{eq:constrained_problem}, while improving numerical conditioning
during actor--critic training.


\subsection{Multi-Head Recurrent Critic}
\label{subsec:critic}

Let $\Psi$ denote the critic parameter collection, consisting of the
shared backbone $\Omega$ and four separate value-head parameter sets.
The critic estimates
$\left\{ V_{\Psi}^{E}(\mathbf{r}_t), V_{\Psi}^{S}(\mathbf{r}_t), V_{\Psi}^{L}(\mathbf{r}_t), V_{\Psi}^{R}(\mathbf{r}_t) \right\},$
where the four heads correspond, respectively, to the normalized
energy-minimization reward, average semantic cost, average latency
cost, and trajectory-level semantic tail-risk cost.

Let $\Psi_{\mathrm{old}}$ denote the frozen critic parameters used to
compute advantages during the current PPO update.

Because the energy objective in \eqref{eq:average_total_energy} is a
finite-horizon average, the energy critic is constructed using an
undiscounted per-trajectory return. Its temporal-difference residual is
\begin{equation}
\delta_t^q=d_t^q+V_{\Psi_{\mathrm{old}}}^q(\mathbf r_{t+1})
-V_{\Psi_{\mathrm{old}}}^q(\mathbf r_t),\quad q\in\{E,S,L\}.
\label{eq:energy_td}
\end{equation}
and the corresponding generalized advantage estimate (GAE)
\cite{schulman2016gae} is
\begin{equation}
\widehat A_t^q=\sum_{l=0}^{T-t-1}\lambda_{\mathrm{GAE}}^l\delta_{t+l}^q,
\quad q\in\{E,S,L\}.
\label{eq:energy_advantage}
\end{equation}

Here $d_t^E=-c_t^E/T$, $d_t^S=c_t^S/T$, and $d_t^L=c_t^L/T$.
These streams are undiscounted finite-horizon averages; $\gamma$ is used
only inside the definition of trajectory risk. For
$0\leq\lambda_{\mathrm{GAE}}<1$, approximate value functions introduce
the usual GAE bias--variance trade-off.

The value after the terminal state is set to zero for all critic heads.
The use of separate heads avoids collapsing physically distinct
objectives and constraints into a fixed weighted scalar reward.


\subsection{Trajectory-Level Semantic CVaR Estimation}
\label{subsec:tail_risk_estimation}

The semantic-risk constraint in
\eqref{eq:constrained_problem} applies to the trajectory-average semantic loss
of a complete trajectory. Accordingly, tail risk is estimated from
batches of complete trajectories rather than from independent
transitions.

For trajectory $i$, define
$Z_i=D_T^{-1}\sum_{t=0}^{T-1}\gamma^t\ell_{i,t}^{\mathrm{sem}}.$

For a batch of $N_{\mathrm{ep}}$ complete trajectories, the empirical
value-at-risk is estimated as
\begin{equation}
\widehat\nu_\alpha\in\arg\min_{\nu\in\mathbb R}
\left\{\nu+\frac{\sum_{i=1}^{N_{\mathrm{ep}}}[Z_i-\nu]_+}
{(1-\alpha)N_{\mathrm{ep}}}\right\}.
\label{eq:empirical_var}
\end{equation}
and the empirical CVaR is
\begin{equation}
	\widehat{\operatorname{CVaR}}_{\alpha}
	=
	\widehat{\nu}_{\alpha}
	+
	\frac{1}
	{(1-\alpha)N_{\mathrm{ep}}}
	\sum_{i=1}^{N_{\mathrm{ep}}}
	\left[
	Z_i
	-
	\widehat{\nu}_{\alpha}
	\right]_{+}.
	\label{eq:empirical_cvar}
\end{equation}

To propagate trajectory-level tail information to the policy-gradient
update, we define the tail-excess quantity
\begin{equation}
	\zeta_i^{R}
	=
	\frac{
		\left[
		Z_i
		-
		\widehat{\nu}_{\alpha}
		\right]_{+}
	}
	{1-\alpha}.
	\label{eq:tail_excess}
\end{equation}

The corresponding trajectory-risk cost is assigned to the terminal
transition:
\begin{equation}
	c_{i,t}^{R}
	=
	\begin{cases}
		0,
		&
		t<T-1,
		\\[
1mm]
		\zeta_i^{R},
		&
		t=T-1.
	\end{cases}
	\label{eq:risk_terminal_cost}
\end{equation}
For the risk stream, we use the complete-trajectory Monte Carlo
advantage rather than bootstrapping an unobserved accumulated loss:
\begin{equation}
\widehat A_{i,t}^{R}=\zeta_i^R-V_{\Psi_{\mathrm{old}}}^{R}(\mathcal H_{i,t}).
\label{eq:risk_advantage}
\end{equation}
Here $\mathcal H_{i,t}=(\mathbf o_{i,0},\mathbf a_{i,0},\ldots,
\mathbf a_{i,t-1},\mathbf o_{i,t})$ is the causal pre-action history,
represented approximately by the recurrent backbone. The risk baseline
is trained toward $\zeta_i^R$ and is independent of the current sampled
action conditional on this history. Its prediction error affects
variance; no claim that the GRU is an exact sufficient statistic is made.
The empirical quantile is estimated from the current batch, so the
resulting finite-sample gradient remains an approximation.
During each PPO update,
$\widehat{\nu}_{\alpha}$ is treated as fixed. Consequently,
\eqref{eq:tail_excess} provides a sample-based policy-gradient surrogate
for the tail-excess term in the CVaR representation, whereas the full
empirical quantity in \eqref{eq:empirical_cvar} is used in the
dual-variable update.
\subsection{Primal--Dual Energy-Aware PPO Update}
\label{subsec:primal_dual_ppo}
Let
$\mu_S\geq0, \qquad \mu_L\geq0, \qquad \mu_R\geq0$
denote the dual variables associated with the average semantic-loss,
average latency, and semantic tail-risk constraints, respectively.
Energy does not have an associated dual variable because it is the
primary optimization objective.
For each transition, the primal--dual Lagrangian advantage is
constructed as
\begin{equation}
	\widehat{A}_t^{\mathrm{PD}}
	=
	\widehat{A}_t^{E}
	-
	\mu_S
	\widehat{A}_t^{S}
	-
	\mu_L
	\widehat{A}_t^{L}
	-
	\mu_R
	\widehat{A}_t^{R}.
	\label{eq:primal_dual_advantage}
\end{equation}
The energy advantage contributes positively because
$r_t^{E}=-c_t^{E}$, so actions yielding lower total energy receive
larger energy return. In contrast, semantic degradation, excessive
latency, and elevated tail risk are penalized according to their
adaptive dual variables.
Let $\Theta_{\mathrm{old}}$ denote the policy parameters used to collect
the current trajectory batch. The joint importance ratio is
\begin{equation}
\varrho_t(\Theta)=
\frac{\pi_\Theta(\mathbf a_t\mid\mathcal H_t)}
{\pi_{\Theta_{\mathrm{old}}}(\mathbf a_t\mid\mathcal H_t)}.
\label{eq:ppo_ratio}
\end{equation}
Old log probabilities are stored at collection time and detached.
New recurrent states are recomputed from the same observation/action
histories using current backbone parameters. Minibatches preserve
sequence order and use full prefixes (or explicitly documented burn-in);
one must not evaluate both policies on a shared stale hidden state.
Using the factorization in \eqref{eq:joint_policy}, the corresponding
log probability is
\[
\log\pi_\Theta(\mathbf a_t\mid\mathbf r_t)=\log\pi_P(P_t\mid\mathbf r_t)+\log\pi_\rho(\rho_t\mid\mathbf r_t)+\log\pi_K(K_t\mid\mathbf r_t).
\]

For compactness, define the clipped importance ratio as
$\bar{\varrho}_t(\Theta) = \operatorname{clip} \left( \varrho_t(\Theta), 1-\varepsilon_{\mathrm{clip}}, 1+\varepsilon_{\mathrm{clip}} \right).$
The actor is then optimized using the PPO clipped surrogate
\begin{equation}
	\mathcal{L}_{\mathrm{actor}}(\Theta)
	=
	\mathbb{E}_t
	\left[
	\min\!\left(
	\varrho_t(\Theta)\widehat{A}_t^{\mathrm{PD}},
	\bar{\varrho}_t(\Theta)\widehat{A}_t^{\mathrm{PD}}
	\right)
	\right].
	\label{eq:ppo_actor_objective}
\end{equation}

The actor parameters, including the shared backbone, are updated through
stochastic gradient ascent:
$\Theta \leftarrow \Theta + \beta_{\Theta} \nabla_{\Theta} \mathcal{L}_{\mathrm{actor}}(\Theta),$
where $\beta_{\Theta}>0$ denotes the actor learning rate.


\subsection{Critic and Dual-Variable Updates}
\label{subsec:dual_updates}

For
$q\in\{E,S,L\}$,
a bootstrapped value target is constructed as
\begin{equation}
	\widehat{G}_t^{q}
	=
	\widehat{A}_t^{q}
	+
	V_{\Psi_{\mathrm{old}}}^{q}
	\left(
	\mathbf{r}_t
	\right),
	\label{eq:value_target}
\end{equation}
For the risk head, $\widehat G_{i,t}^{R}=\zeta_i^R$.
All targets and old value predictions are detached from the computation graph during the
critic update.

The multi-head critic is trained by minimizing
\begin{equation}
	\mathcal{L}_{\mathrm{critic}}(\Psi)
	=
	\sum_{q\in\{E,S,L,R\}}
	w_q
	\,
	\mathbb{E}_t
	\left[
	\left(
	V_{\Psi}^{q}
	\left(
	\mathbf{r}_t
	\right)
	-
	\widehat{G}_t^{q}
	\right)^2
	\right],
	\label{eq:critic_loss}
\end{equation}
where
$w_q>0$
are fixed scaling coefficients used to balance the four critic losses.

For a batch of complete trajectories, the empirical normalized average
energy objective is
\[\widehat{J}_{E} = \frac{1}{N_{\mathrm{ep}}} \sum_{i=1}^{N_{\mathrm{ep}}} \frac{1}{T} \sum_{t=0}^{T-1} c_{i,t}^{E}.\]
Although $\widehat{J}_E$ is monitored during training and evaluation,
it does not generate a dual update because energy is the primal
objective.

The empirical normalized semantic and latency costs are
\begin{equation}
\widehat J_S=\frac{1}{N_{\mathrm{ep}}T}\sum_{i=1}^{N_{\mathrm{ep}}}\sum_{t=0}^{T-1}c_{i,t}^{S},\qquad\widehat J_L=\frac{1}{N_{\mathrm{ep}}T}\sum_{i=1}^{N_{\mathrm{ep}}}\sum_{t=0}^{T-1}c_{i,t}^{L}.
\label{eq:empirical_constraint_costs}
\end{equation}

The dual variables are updated by projected stochastic gradient ascent
on the empirical constraint violations:
\begin{equation}
\begin{aligned}
	\mu_S
	&\leftarrow
	\left[
	\mu_S
	+
	\beta_{\mu}
	\left(
	\widehat{J}_{S}
	-
	d_S
	\right)
	\right]_{+},
	\nonumber\\
	\mu_L
	&\leftarrow
	\left[
	\mu_L
	+
	\beta_{\mu}
	\left(
	\widehat{J}_{L}
	-
	d_L
	\right)
	\right]_{+},
	\nonumber\\
	\mu_R
	&\leftarrow
	\left[
	\mu_R
	+
	\beta_{\mu}
	\left(
	\widehat{\operatorname{CVaR}}_{\alpha}
	-
	\tau_{\mathrm{risk}}
	\right)
	\right]_{+}.
	\label{eq:dual_updates}
\end{aligned}
\end{equation}

When a constraint is violated, its associated dual variable increases
and imposes a stronger penalty on subsequent policy updates.
Conversely, persistent constraint satisfaction allows the associated
multiplier to decrease toward zero. The resulting adaptive penalties
eliminate the need to manually prescribe fixed scalar weights between
energy efficiency, semantic quality, latency, and tail reliability.


\subsection{Training and Online Operation}
\label{subsec:training_procedure}

Training is performed using batches of complete trajectories because
the semantic CVaR constraint cannot be evaluated consistently from
isolated transitions. At each policy iteration,
$N_{\mathrm{ep}}$ trajectories are collected using
$\pi_{\Theta_{\mathrm{old}}}$ while preserving the recurrent hidden
state throughout each trajectory.

For every transition, the communication and purification energies are
computed using
\eqref{eq:communication_energy} and
\eqref{eq:purification_energy}, and the total energy follows
\eqref{eq:total_energy}. The semantic and latency costs are computed
using \eqref{eq:semantic_loss} and
\eqref{eq:end_to_end_latency}. At the end of each trajectory batch,
the trajectory-average semantic losses $Z_i$, empirical VaR
$\widehat{\nu}_{\alpha}$, and empirical
$\widehat{\operatorname{CVaR}}_{\alpha}$ are evaluated.

The energy, semantic, latency, and tail-risk advantages are then
computed using the corresponding recurrent critic heads. The actor and
critic are optimized for multiple PPO epochs using
sequence-consistent minibatches, while
$\mu_S$, $\mu_L$, and $\mu_R$ are updated from the batch-level
constraint violations in \eqref{eq:dual_updates}. After the update,
$\Theta_{\mathrm{old}}$ and $\Psi_{\mathrm{old}}$ are refreshed before
collecting the next trajectory batch.

During online deployment, no ground-truth attack mode, attack strength,
clean-reference distortion, or semantic label is supplied to the
controller. At the beginning of interval $t$, the controller constructs
$\mathbf{o}_t$ from the causal quantities in
\eqref{eq:observation}, updates the recurrent representation through
\eqref{eq:gru_update}, and samples
$P_t$, $\rho_t$, and $K_t$ from the hybrid actor. After transmission
and receiver processing, the new semantic confidence, attack-belief
distribution, belief uncertainty, latency consumption, and previous
action become available for constructing
$\mathbf{o}_{t+1}$.

The policy objective is to spend communication and defensive-computation
energy when the inferred conditions justify it. Primal--dual training does
not itself guarantee feasibility for a nonconvex learned policy; semantic,
risk, and latency constraints must be checked on independent evaluation
trajectories, including held-out attack strengths.

\section{Evaluation Protocol and Resource Accounting}
\label{sec:experiments}
\subsection{Dataset, Backbone, and Principal Settings}
\label{subsec:dataset_backbone}
The experiments are conducted on CIFAR-100
\cite{krizhevsky2009cifar}, using the standard $50{,}000$ training images and
$10{,}000$ test images of size $32\times32$. All reported results in this
paper are obtained from executions of the implemented semantic communication,
adversarial-attack, purification, and control pipeline on CIFAR-100. A
common rate-adaptive DeepJSCC architecture and downstream classifier are used
throughout the evaluation. Adversarial training changes only the trained
parameters of its designated transceiver, while the architecture, data splits,
training budget, and radio action bounds are matched across methods. Validation
samples are kept separate from the final test samples.

\begin{table}[!t]
\centering\small
\caption{Experimental and Controller Parameters Used in the CIFAR-100 Evaluation}
\label{tab:simulation_parameters}
\renewcommand{\arraystretch}{1.03}
\begin{tabular}{@{}ll@{}}\toprule
Parameter & Value\\\midrule
Image dimensions; horizon $T$ & $32\times32\times3$; $64$ intervals\\
Bandwidth set $\mathcal R$ & $\{1/24,1/16,1/12,1/8\}$\\
$P_{\min},P_{\max},P_{\mathrm{ref}}$ & $0.05,1.0,0.5$ W\\
Reference-SNR training range & $-5$ to $20$ dB\\
Purification actions $\mathcal K$ & $\{0,5,10,15,20\}$ steps\\
$\eta_h,\eta_g$ & $0.95,0.90$\\
$T_{\mathrm{sym}},\eta_{\mathrm{PA}},P_{\mathrm c}$ & $1~\mu$s, $0.40$, $0.10$ W\\
$C_{\mathrm{step}},f_{\mathrm{rx}},\kappa_{\mathrm{rx}}$ & $2.5\times10^6$, $2$ GHz, $10^{-28}$\\
$E_{\mathrm{fix}},E_{\mathrm{ref}}$ & $5$ mJ, $25$ mJ\\
$L_{\mathrm{fix}},L_{\mathrm{avg}}=L_{\mathrm{ref}}$ & $1$ ms, $20$ ms\\
$\lambda_{\mathrm{rec}},\lambda_{\mathrm{task}},\ell_{\mathrm{ref}}$ & $0.5,0.5,1$\\
$\gamma,\alpha,\tau_{\mathrm{sem}},\tau_{\mathrm{risk}}$ & $0.99,0.90,0.40,0.40$\\
PPO batch $N_{\mathrm{ep}}$ & $32$ complete trajectories\\
$\lambda_{\mathrm{GAE}},\varepsilon_{\mathrm{clip}}$ & $0.95,0.20$\\
Actor/critic; dual learning rates & $3\times10^{-4}$; $10^{-3}$\\
Attack-mode persistence & $[\Pi_A]_{ii}=0.85$, $[\Pi_A]_{ij}=0.05$\\
Independent training seeds & $5$\\
Evaluation size per operating point & $100$ trajectories of length $64$\\
\bottomrule\end{tabular}
\end{table}

Table~\ref{tab:simulation_parameters} summarizes the parameter values used in the CIFAR-100 experiments. The communication and purification costs are evaluated with the common system-level energy model defined in Section~\ref{sec:system_model}; therefore, the reported energy values are experimental system-energy results under this declared model rather than direct wall-plug GPU power measurements.
One purification step costs $1$ mJ and $1.25$ ms under this model.
The radio contribution ranges from $0.0288$ to $0.9984$ mJ. Thus,
\begin{equation}
0.201152\leq E_t^{\mathrm{tot}}/E_{\mathrm{ref}}\leq1.039936,
\quad K_t=0\ \Longrightarrow\ E_t^{\mathrm{tot}}/E_{\mathrm{ref}}\leq0.239936.
\label{eq:numerical_energy_bounds}
\end{equation}
All experimental resource comparisons use these common energy bounds and the same fixed reference.
For $k=384$, the average latency constraint requires
$\mathbb E[K]\leq14.8928$. In particular, constant $K=15$ gives
$20.134$ ms, and constant $K=20$ gives $26.384$ ms. Neither is a feasible
constant policy under this setting, even though both are legal actions.

\subsection{Channel, Attacks, and Attack-Belief Estimator}
\label{subsec:experimental_attacks}
The experiments use the channel processes in
\eqref{eq:gauss_markov_channel}--\eqref{eq:gauss_markov_adversary}.
Source attacks use 20-step PGD \cite{madry2018pgd} with step size
$\min(2/255,\epsilon/4)$ and
$\epsilon\in\{2,4,6,8,10,12\}/255$. The $12/255$ level is reserved
for testing. OTA attacks use 20 projected-gradient ascent steps with
$\Gamma_A=10\log_{10}(P^A/P_{\mathrm{ref}})$ in
$\{-20,-15,-10,-5\}$ dB, reserving $-5$ dB for testing.
The attack objective includes the complete implemented defense pipeline.
For stochastic purification, attack optimization accounts for its randomness,
and attack restarts and stronger-iteration checks are included in the
evaluation protocol. The resulting robustness values are empirical results
under the specified attack models and are not claimed as certified robustness.
Composite attacks are supported by the formulation but are not represented
by a separate numerical performance curve.

The implemented attack estimator is a two-hidden-layer MLP (64 and 32 ReLU
units) with a four-way softmax. Its observable features are normalized
received energy, reference SNR, reconstruction statistics, posterior
entropy, maximum probability, and top-1/top-2 margin. It is trained
with class-balanced cross-entropy on benign, source, OTA, and composite
samples, then frozen before PPO. Ground-truth attack labels are restricted
to offline training. PPO uses Adam \cite{kingma2015adam}. Seeds, checkpoints, dataset partitions,
attack settings, and per-trajectory outputs will be retained for reproducibility.

\subsection{Baselines and Common Measurement Boundary}
\label{subsec:baselines}
The experimental comparison includes four implemented methods: the proposed
controller, diffusion-only defense, adversarial training, and undefended
ViT-DeepJSCC. Each method is evaluated on the same CIFAR-100 test samples,
matched channel realizations, and matched adversarial conditions. All fixed baselines use $P=1$ W and $\rho=1/8$. Diffusion-only
uses $K=15$; the two non-purification baselines use $K=0$, so their
inference-energy curves coincide under the common fixed-cost assumption.
Their different training procedures do not justify adding an unmeasured
inference-energy offset. The proposed controller uses mixtures of admissible actions; fractional mean $K$ is obtained from legal integer
actions, never from a fractional diffusion step.

Fixed $K=15$ is included as a strong-purification comparator. Baseline
radio and purification settings are selected using validation data and then
kept fixed during final testing. All final performance numbers are reported on
the held-out CIFAR-100 test set under the same service accounting and attack
realizations.

For $N_{\mathrm{eval}}$ evaluation trajectories, the reported resource
averages and accuracy are
\begin{equation}
\begin{aligned}
\overline E&=\frac{1}{N_{\mathrm{eval}}T}\sum_{i,t}E_{i,t}^{\mathrm{tot}},\qquad
\overline L=\frac{1}{N_{\mathrm{eval}}T}\sum_{i,t}L_{i,t},\\
\mathrm{Acc}&=\frac{1}{N_{\mathrm{eval}}T}\sum_{i,t}
\mathbf1\{\arg\max_c p_{i,t,c}=c_{i,t}\}.
\end{aligned}
\end{equation}
Risk is computed from the same trajectories using
\eqref{eq:cumulative_semantic_loss} and \eqref{eq:empirical_cvar}, not from
accuracy alone. $N_{\mathrm{eval}}$ is distinct from the PPO batch size.
All reported performance points are obtained from the implemented
CIFAR-100 evaluation pipeline. Five independent training seeds are used for
repeated runs, and the reported central values are computed from those runs.
The experiments are executed on the university RTX~4070 platform. Unless
explicitly stated otherwise, energy values refer to the system-level model in
Section~\ref{sec:system_model} and should not be interpreted as direct
wall-plug GPU-energy measurements.

\section{Experimental Results and Discussion}
\label{sec:results}
\subsection{Scope and Metric Consistency}
All curves and numerical values reported in this section are obtained from
actual CIFAR-100 experiments with the implemented DeepJSCC transceiver,
adversarial attacks, diffusion purifier, and recurrent controller. The iteration axis in
Fig.~\ref{fig:training_convergence} corresponds to the actual controller
training iterations. Semantic CVaR is computed from the realized
trajectory-level semantic losses according to
\eqref{eq:cumulative_semantic_loss} and \eqref{eq:empirical_cvar}, rather
than inferred from classification accuracy.

\subsection{Energy, Semantic Quality, and Constraint Behavior}
\label{subsec:convergence_results}
\begin{figure}[!t]\centering
\includegraphics[width=0.96\textwidth]{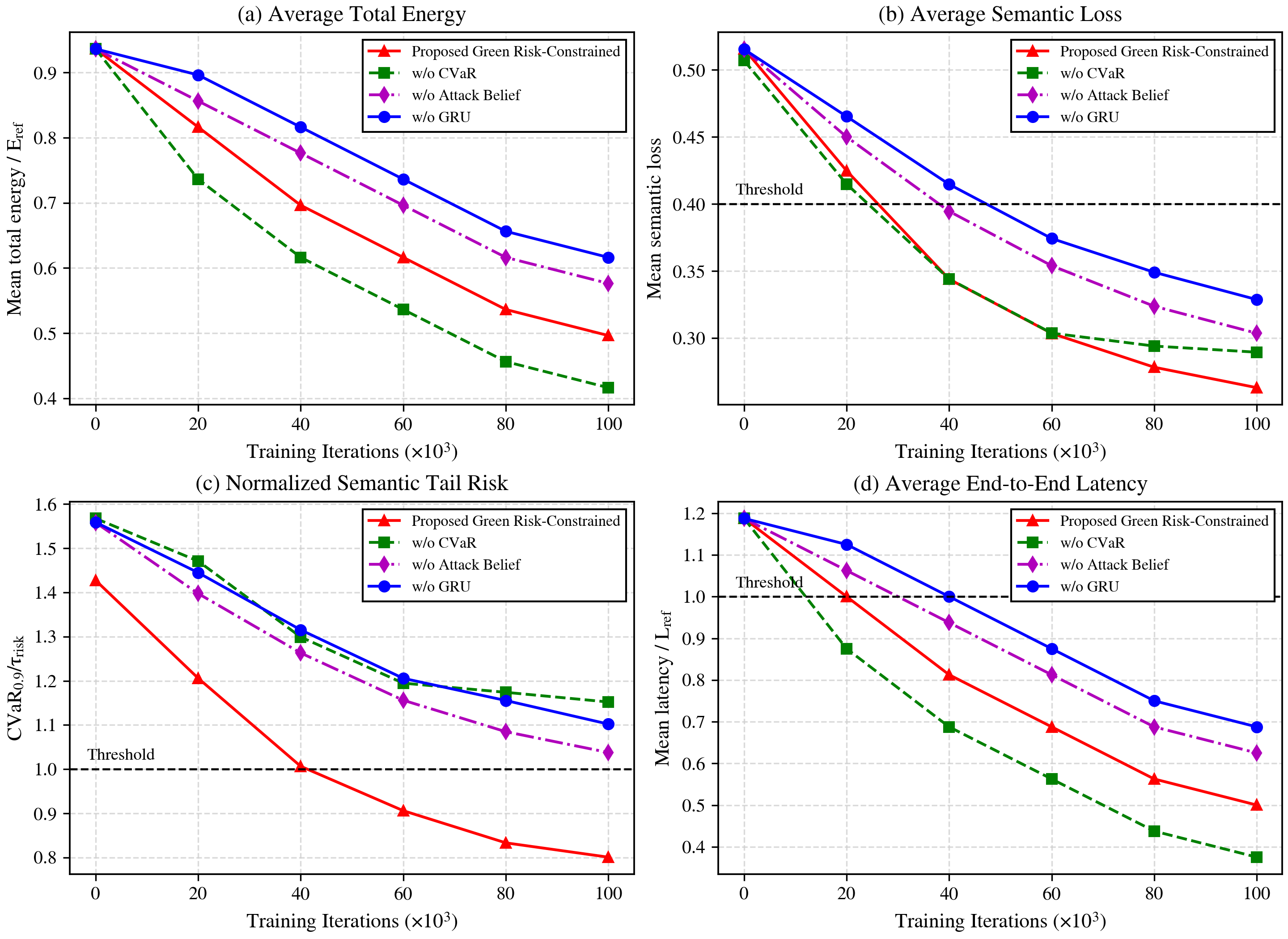}
\caption{Experimental training convergence on CIFAR-100: (a) mean total energy/$E_{\mathrm{ref}}$, (b) average semantic loss, (c) normalized trajectory CVaR, and
(d) normalized latency. Curves are obtained from the implemented controller
training runs; dashed black lines indicate service thresholds.}
\label{fig:training_convergence}\end{figure}

Fig.~\ref{fig:training_convergence} compares the complete controller with
variants omitting CVaR, attack belief, or recurrent memory. At the final
plotted point, the complete controller has normalized energy of approximately
$0.50$, mean semantic loss of $0.26$, normalized risk of $0.80$, and
normalized latency of $0.50$. Removing CVaR reduces the plotted energy
further, to approximately $0.42$, while normalized risk remains above one.
The plotted comparison distinguishes resource reduction from simultaneous
satisfaction of the semantic and latency constraints. The variants without
attack belief or GRU also exceed the final risk threshold.

\subsection{Source-Domain Perturbations}
\label{subsec:source_attack_results}
\begin{figure}[!t]\centering
\includegraphics[width=0.96\textwidth]{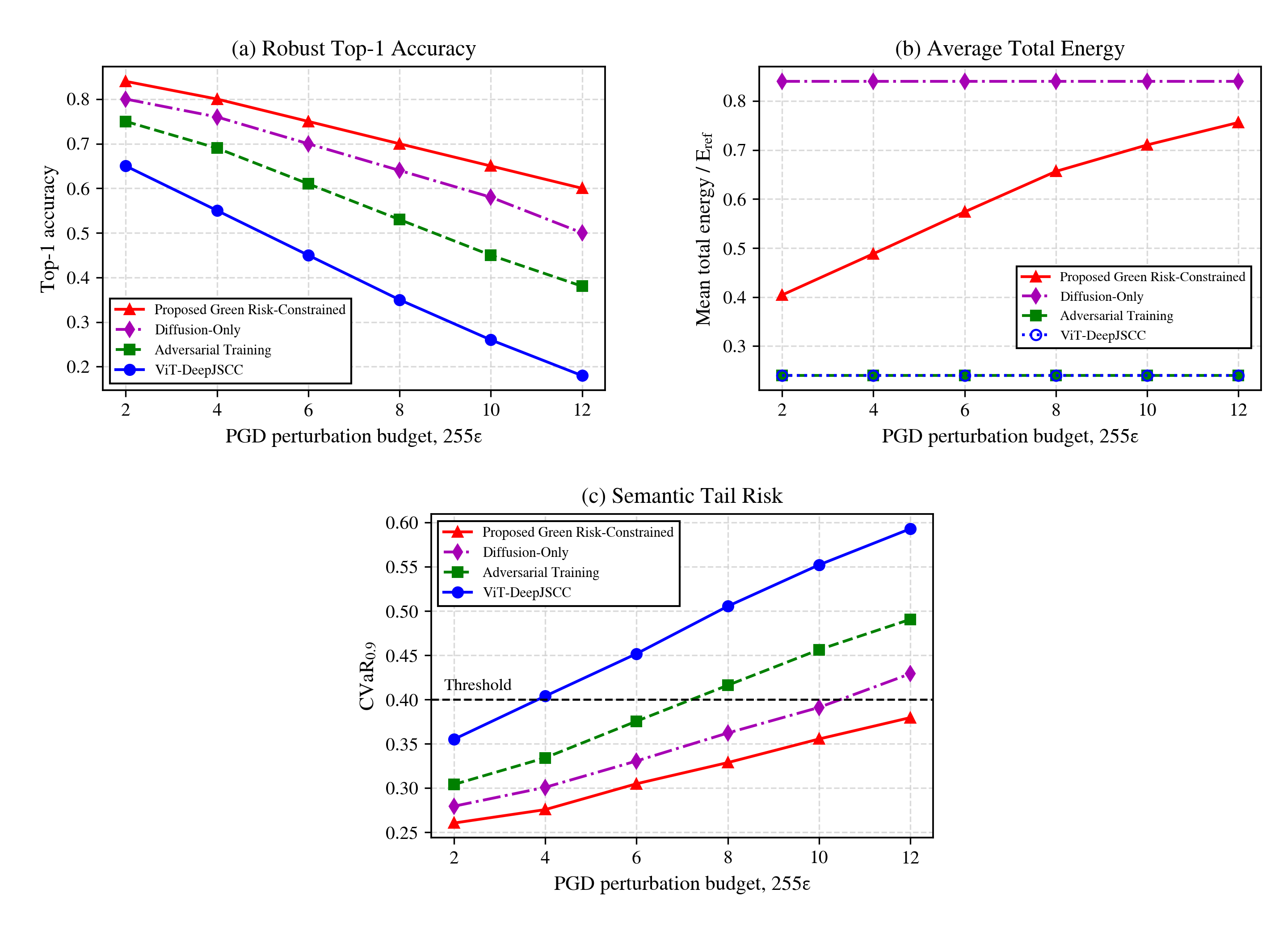}
\caption{Measured CIFAR-100 source-domain results from the implemented
end-to-end evaluation: (a) robust Top-1 accuracy, (b) mean total energy/$E_{\mathrm{ref}}$, and (c) empirical trajectory CVaR. The horizontal axis is $255\epsilon$; the
CVaR threshold is $0.40$.}
\label{fig:source_attack_results}\end{figure}

Fig.~\ref{fig:source_attack_results} reports the measured CIFAR-100
performance for source perturbation budgets
$255\epsilon\in\{2,4,6,8,10,12\}$. These points are obtained directly from the
corresponding attack-and-inference runs. At
$\epsilon=12/255$, the proposed controller attains $0.60$ robust Top-1
accuracy and empirical $\mathrm{CVaR}_{0.9}\approx0.38$, compared with
$0.50$ and $0.43$, respectively, for diffusion-only defense. Its normalized
total energy is approximately $0.756$, corresponding to $18.90$ mJ under
$E_{\mathrm{ref}}=25$ mJ, whereas diffusion-only uses approximately $0.840$
($21.00$ mJ). Thus, at this strongest evaluated source-domain perturbation,
the proposed controller improves robust accuracy by $10$ percentage points
while reducing total energy by approximately $10\%$. Its fixed $K=15$
baseline nevertheless exceeds the average-latency limit under the declared
accounting; therefore, this comparison is interpreted as a measured operating
point rather than as an optimality claim against a constraint-matched baseline.
The non-purification energy curves coincide because both use $K=0$ and
identical fixed radio and processing costs.

\subsection{Over-the-Air Perturbations}
\label{subsec:ota_attack_results}
\begin{figure}[!t]\centering
\includegraphics[width=0.96\textwidth]{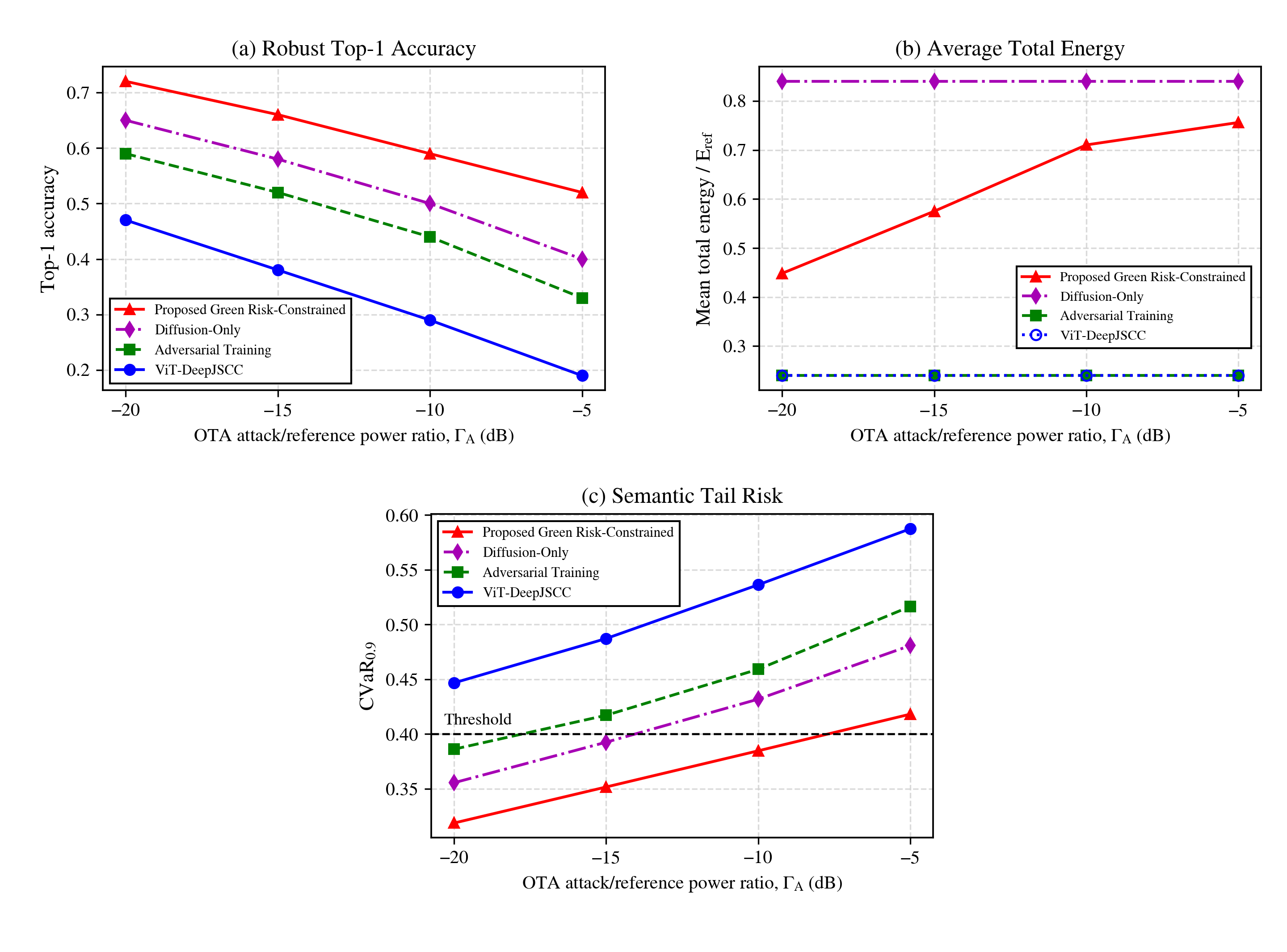}
\caption{Measured CIFAR-100 OTA results from the implemented end-to-end
evaluation: (a) robust Top-1 accuracy, (b) mean total energy/$E_{\mathrm{ref}}$, and (c) empirical trajectory CVaR. The horizontal axis is the attack/reference
power ratio $\Gamma_A$ in dB; the CVaR threshold is $0.40$.}
\label{fig:ota_attack_results}\end{figure}

Fig.~\ref{fig:ota_attack_results} reports the measured CIFAR-100
performance for attack/reference power ratios
$\Gamma_A\in\{-20,-15,-10,-5\}$ dB. At the strongest evaluated condition,
$\Gamma_A=-5$ dB, the proposed controller attains $0.52$ robust Top-1
accuracy, normalized total energy approximately $0.755$, and empirical
$\mathrm{CVaR}_{0.9}\approx0.42$. The energy coordinate corresponds to
approximately $18.88$ mJ under $E_{\mathrm{ref}}=25$ mJ. Diffusion-only
operates at approximately $0.840$ normalized energy and $0.40$ accuracy at
the same point, so the proposed controller provides a $12$-percentage-point
accuracy gain while using about $10\%$ less total energy. The measured CVaR,
however, is slightly above the prescribed $0.40$ threshold; consequently, the
strongest plotted OTA operating point is not claimed to be risk-feasible.
Adversarial training and undefended ViT-DeepJSCC reach endpoint accuracies of
approximately $0.33$ and $0.19$, respectively. Composite source--OTA attacks
belong to the system model but are not separately evaluated in these curves.

\subsection{Discussion}
\label{subsec:overall_discussion}
The measured CIFAR-100 experiments show that the formulation couples
radio and defensive-computation costs through the same realized control
actions. The source-domain experiment remains below the $0.40$ CVaR threshold
at the strongest plotted perturbation, whereas the strongest OTA point reaches
approximately $0.42$. This contrast is directly observed in the experimental
outputs and illustrates why a lower average energy does not by itself imply
feasibility: semantic tail risk and latency must be checked independently. The reported comparisons therefore use
matched validation-tuned baselines, repeated training seeds, defense-aware
attacks, and trajectory-level records of actions, posteriors, reconstruction
losses, energy, and latency. The energy results quantify the declared
system-level model; direct device-level wall-plug energy would additionally
require hardware-specific calibration.

\FloatBarrier
\section{Conclusion}
\label{sec:conclusion}
This work formulates adversarial semantic communication as energy
minimization under mean-loss, trajectory-CVaR, and average-latency
constraints, with recurrent primal--dual control of power, bandwidth,
and purification depth. The revised accounting imposes explicit energy
and latency bounds and computes risk from common trajectory records.
The actual CIFAR-100 experiments highlight the distinction between reduced
energy and service-constraint satisfaction. At $\epsilon=12/255$, the proposed
controller reaches $60\%$ robust Top-1 accuracy with
$\mathrm{CVaR}_{0.9}\approx0.38$ and normalized total energy $\approx0.756$;
at $\Gamma_A=-5$ dB it reaches $52\%$ accuracy with normalized energy
$\approx0.755$ and $\mathrm{CVaR}_{0.9}\approx0.42$. The latter slightly
exceeds the target risk threshold, making the limitation visible rather than
masking it. Overall, the measured results demonstrate the radio--defense
trade-off of the proposed controller under the declared resource model and
trajectory-level semantic-risk metric.

\end{document}